\documentclass[conference]{IEEEtran}
\usepackage[numbers,sort&compress]{natbib}
\usepackage{graphicx}
\usepackage{dirtytalk}
\usepackage{etoolbox}
\usepackage{booktabs}
\usepackage{multirow}

\usepackage[mathscr]{euscript}

\usepackage{caption}
\usepackage{subcaption}

\usepackage[table,xcdraw]{xcolor}
\makeatletter

\patchcmd{\@makecaption}
 {\\}
 {.\ }
 {}
 {}
 
\ifCLASSINFOpdf
\else
\fi
\usepackage{amsmath}
\usepackage{amssymb}
\usepackage{threeparttable}
\usepackage{scalerel}
\usepackage{tikz}
\usetikzlibrary{svg.path}

\definecolor{orcidlogocol}{HTML}{A6CE39}
\tikzset{
    orcidlogo/.pic={
        \fill[orcidlogocol] svg{M256,128c0,70.7-57.3,128-128,128C57.3,256,0,198.7,0,128C0,57.3,57.3,0,128,0C198.7,0,256,57.3,256,128z};
        \fill[white] svg{M86.3,186.2H70.9V79.1h15.4v48.4V186.2z}
        svg{M108.9,79.1h41.6c39.6,0,57,28.3,57,53.6c0,27.5-21.5,53.6-56.8,53.6h-41.8V79.1z M124.3,172.4h24.5c34.9,0,42.9-26.5,42.9-39.7c0-21.5-13.7-39.7-43.7-39.7h-23.7V172.4z}
        svg{M88.7,56.8c0,5.5-4.5,10.1-10.1,10.1c-5.6,0-10.1-4.6-10.1-10.1c0-5.6,4.5-10.1,10.1-10.1C84.2,46.7,88.7,51.3,88.7,56.8z};
    }
}

\newcommand\orcidicon[1]{\href{https://orcid.org/#1}{\mbox{\scalerel*{
                \begin{tikzpicture}[yscale=-1,transform shape]
                \pic{orcidlogo};
                \end{tikzpicture}
            }{|}}}}

\newcommand{\figref}[1]{Fig.\hspace{1mm}\ref{#1}}
\newcommand{\tabref}[1]{Table\hspace{1mm}\ref{#1}}
\newcommand{\equref}[1]{Eq \hspace{1mm}(\ref{#1})}

\usepackage{hyperref}

\begin{document}

\title{A Practical Framework for Unsupervised Structure Preservation Medical Image Enhancement}
% author names and affiliations
% use a multiple column layout for up to three different
% affiliations

\author{\IEEEauthorblockN{Quan Huu Cap\IEEEauthorrefmark{1}\IEEEauthorrefmark{2}, Atsushi Fukuda\IEEEauthorrefmark{2}, and Hitoshi Iyatomi\IEEEauthorrefmark{1}}
\IEEEauthorblockA{quan.cap@aillis.jp\quad atsushi.fukuda@aillis.jp\quad iyatomi@hosei.ac.jp}
\IEEEauthorblockA{\IEEEauthorrefmark{1}AI Development Department, Aillis, Inc., Tokyo, Japan}
\IEEEauthorblockA{\IEEEauthorrefmark{2}Applied Informatics, Graduate School of Science and Engineering, Hosei University, Tokyo, Japan}
}

% make the title area
\maketitle
% As a general rule, do not put math, special symbols or citations
% in the abstract
% ABSTRACT
\begin{abstract}
    Medical images are extremely valuable for supporting medical diagnoses. 
However, in practice, low-quality (LQ) medical images, such as images that are hazy/blurry, have uneven illumination, or are out of focus, among others, are often obtained during data acquisition. 
This leads to difficulties in the screening and diagnosis of medical diseases. 
Several generative adversarial networks (GAN)-based image enhancement methods have been proposed and have shown promising results. 
However, there is a quality-originality trade-off among these methods in the sense that they produce visually pleasing results but lose the ability to preserve originality, especially the structural inputs. 
Moreover, to our knowledge, there is no objective metric in evaluating the structure preservation of medical image enhancement methods in unsupervised settings due to the unavailability of paired ground-truth data. 
In this study, we propose a framework for practical unsupervised medical image enhancement that includes (1) a non-reference objective evaluation of structure preservation for medical image enhancement tasks called Laplacian structural similarity index measure (LaSSIM), which is based on SSIM and the Laplacian pyramid, and (2) a novel unsupervised GAN-based method called Laplacian medical image enhancement (LaMEGAN) to support the improvement of both originality and quality from LQ images. The LaSSIM metric does not require clean reference images and has been shown to be superior to SSIM in capturing image structural changes under image degradations, such as strong blurring on different datasets. 
The experiments demonstrated that our LaMEGAN achieves a satisfactory balance between quality and originality, with robust structure preservation performance while generating compelling visual results with very high image quality scores. 
The code will be made available at \url{https://github.com/AillisInc/USPMIE}. 
\end{abstract}

% KEYWORDS

% For peer-review papers, this IEEEtran command inserts a page break and
% creates the second title. It will be ignored for other modes.
% INTRODUCTION
\section{Introduction}
    According to the World Health Organization (WHO), oral diseases are among the most prevalent ailments globally, with significant health, social, and economic impacts \cite{who2022oral}. 
More than 3.5 billion people suffer from oral diseases, without any notable improvement between 1990 and 2017 \cite{james18}. 
In 2015, oral diseases accounted for 357 billion USD in direct costs and 188 billion USD in indirect costs worldwide \cite{righolt18}. 
Therefore, reducing these burdens is essential. 
Throat images are a rich source of valuable visual information and are very effective for helping clinicians in making the correct detection, diagnosis, and treatment of oral diseases. 
Recently, throat images have been highlighted as useful for the diagnosis of influenza \cite{okiyama22} and the detection of severe pharyngitis \cite{yoo20}. 
Thus, throat image analysis has become even more important.  
% FIGURE 1
\begin{figure}[!t]
\centering
\includegraphics[width=0.99\linewidth]{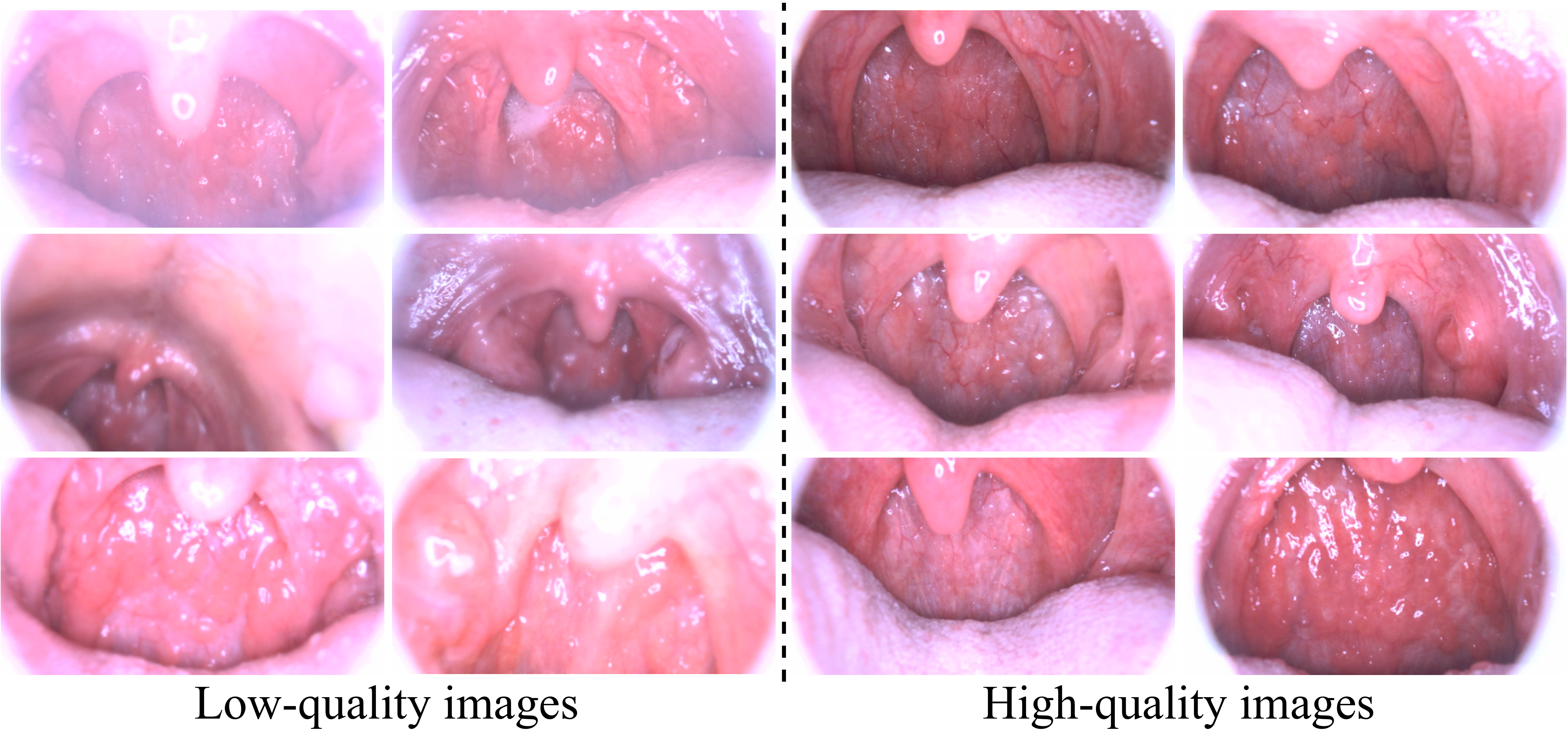}
\caption{
    Comparison of low-quality (hazy, out-of-focus, overexposure) and high-quality throat images (clean, natural).
}
\label{fig:fig_1}
\end{figure}

However, during the acquisition process, the quality of the throat images can easily be compromised due to the shooting conditions. 
Low-quality (LQ) throat images typically appear hazy/blurry, out of focus, with uneven illumination, and so on. 
\figref{fig:fig_1} shows examples of actual throat images with unsatisfactory quality (left) compared to the ones with high quality (HQ) (right), as determined by physicians specializing in otolaryngology. 
These LQ images can greatly impair the diagnosis; hence, image enhancement methods to improve the quality of medical images are in great demand.  

Classic image enhancement techniques, such as adaptive histogram equalization (AHE) \cite{pizer87}, genetic algorithm \cite{holland92}, dark prior channel \cite{he10darkchannel}, Krill herd heuristic optimization \cite{gandomi12}, and weighted guided filtering \cite{li14weighted}, have been applied to improve the quality of medical images. 
However, these methods are still handcrafted and need to be carefully designed.  

Over the past decade, convolutional neural networks (CNNs) have demonstrated their power in image restoration tasks \cite{dong20multi, orest19deblurganv2, qin20ffa, yiqun21, liang21swinir, wu21contrastive,wang22self,mou22deep,zamir22restormer,lu22transformer}. 
Despite the successes, most of these CNNs still require a large number of LQ/HQ image pairs in the training phase, which are very difficult to obtain in the medical field. 
To solve this issue, several methods that do not require paired training data have been proposed for image enhancement \cite{dudhane19cdnet, golts19unsupervised, jin20unsupervised, li21you, an21unsupervised, chen22unpaired, li22usid, wang22cycle}. 
Among these unpaired methods, CycleGAN \cite{zhu2017unpaired} and its successors are well used, thanks to the introduction of the cycle-consistency loss which assumes that the generated image can be transformed back to its original input form. 
Inspired by the practicality of CycleGAN, several medical image enhancement methods have also been proposed \cite{zhao19data, luo20dehaze, wan22rentinal, chen22novel, ma21structure}. 
Although these methods are recognized as very practical and effective, they still have some limitations.  

\emph{First}, these methods only consider evaluating the quality aspect of the enhanced outputs by applying the non-reference image quality assessment (NR-IQA) and/or by subjective visual comparison. 
For medical image enhancement evaluation, it is crucial to also take into consideration the aspect of originality preservation (e.g., structure, texture, color) from the inputs. 
In this paper, structure preservation is the main aspect that will be addressed. The structural similarity index measure (SSIM) \cite{wang04ssim}, a measure of image similarity using simple image statistics, is commonly used in general image evaluation and has been used in medical applications to measure the structural features between LQ input and enhanced output images \cite{rundo19medga, acharya21, zhao2019new, subramani19fuzzy, li19medical, kandhway20novel, wang22lung, zhang22medical}. 
However, this metric can produce inaccurate results if the input images are in poor condition. 
In addition, using SSIM as a metric for medical image evaluation is not recommended if no \say{perfect} reference images are available \cite{chow16review}. 
To our knowledge, no objective evaluation metric for evaluating structure preservation in the absence of reference images has been reported.  

\emph{Second}, unsupervised GAN-based image-to-image enhancement methods have a quality-originality trade-off because they generate HQ images but lose structure preservation from the inputs \cite{cohen18distribution, wagner21structure}. 
On the contrary, the image quality decreases if the ability to preserve the structure is increased, which is due to the unavailability of reference images. 
Thus, as the structure preservation is increased, noise or hazy information from the input LQ image may remain. 
For this reason, developing a method to balance this trade-off is essential.  

To address the abovementioned issues, we propose a framework consisting of two methods: (1) a new non-reference structure preservation evaluation for medical image enhancement tasks, and (2) an unsupervised medical image enhancement method that balances the trade-off between image quality and structure preservation. 
The first proposed non-reference evaluation of structure preservation is the Laplacian structural similarity index measure (LaSSIM). 
We show that SSIM cannot be properly used for evaluation in situations of image degradation, such as high blur, whereas the Laplacian pyramid technique (LP) \cite{burt87laplacian} is an effective method to represent the global structure of an image under these circumstances. 
Based on this knowledge, our LaSSIM applies SSIM in the LP space to measure the structural similarity between two images. 
The LaSSIM is resistant to degradation, such as blurring, that often occurs in medical images and is able to capture structural changes without the need for clean reference images, which is practical in real-world settings. 
Along with this evaluation approach, we introduce an unsupervised medical image enhancement method called Laplacian medical image enhancement (LaMEGAN). 
The LaMEGAN improves the image quality while preserving the important original structure of the involved region to support clinicians in making medical detection and diagnosis. 
To preserve the structure from the LQ inputs, we introduce optimized perceptual loss \cite{johnson16perceptual} in the LP space. 
We demonstrate that this constraint helps reduce the artifacts and produce perceptually pleasing results while preserving the structure information from the inputs. 
The proposed LaMEGAN can achieve higher performance compared with other unsupervised image-to-image translation approaches in the throat image enhancement task. 
\vspace{\baselineskip}

\section{Related Works}
    % FIGURE 2
\begin{figure*}[!t]
\centering
\includegraphics[width=0.8\linewidth]{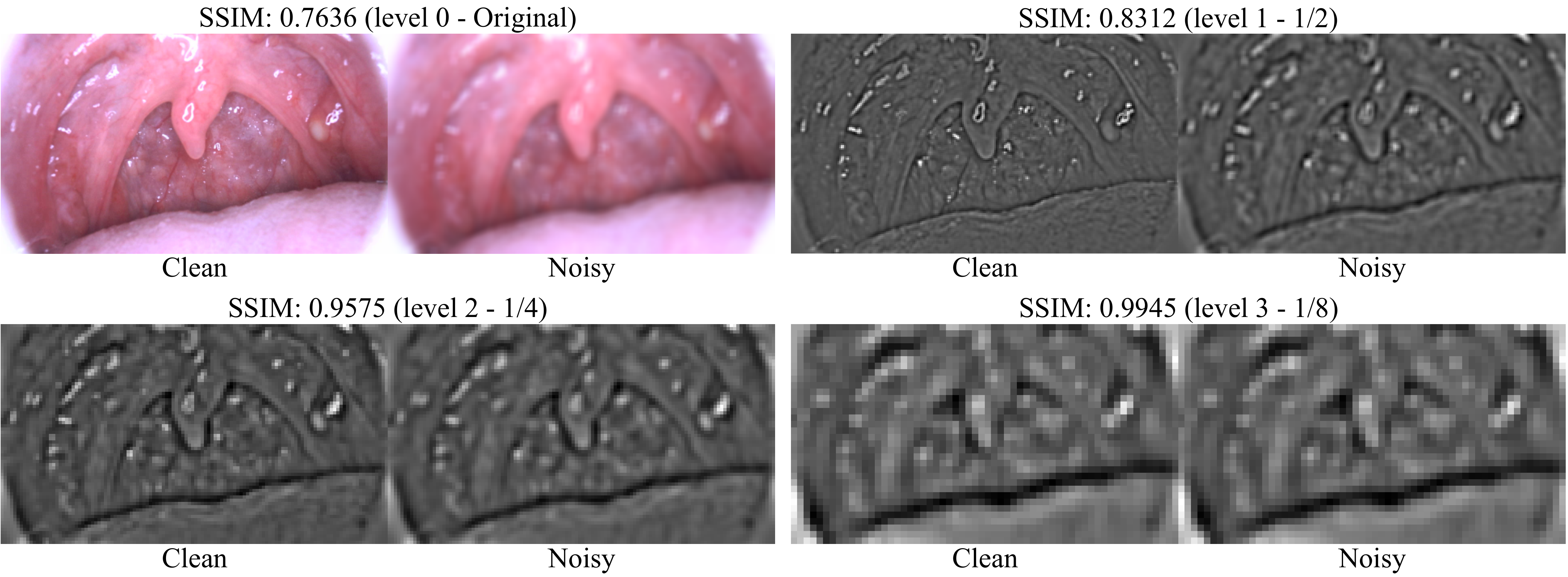}
\caption{
    Comparison of SSIM values on original color images and different LP levels. At LP level 3, the underlying structure of the two images are very close.
}
\label{fig:fig_2}
\end{figure*}

\subsection{Medical image enhancement}
\vspace{\baselineskip}
Several classic image enhancement techniques, such as AHE \cite{pizer87}, genetic algorithm \cite{holland92}, dark prior channel \cite{he10darkchannel}, Krill herd heuristic optimization \cite{gandomi12}, and weighted guided filtering \cite{li14weighted}, have been utilized to improve the quality of medical images. 
These techniques cover almost all imaging modalities, including magnetic resonance imaging \cite{rundo19medga, subramani19fuzzy, kandhway20novel, acharya21}, computed tomography \cite{li19medical, kandhway20novel, wang22lung}, X-ray \cite{rui17medical, gong19soft, kandhway20novel}, and endoscopic images \cite{fang21color, zhang22medical}. 
Despite achieving promising results, these classic methods require manual and empirical tuning, which are tedious processes. 
With the wide variety and complex appearance of LQ color medical images, such as strong degradations like haziness/blurriness (see \figref{fig:fig_1}), we argue that these methods will not work well in such cases. 
Thus, the practical results are not yet confirmed and are still left open.  

With the ability to automatically learn meaningful features from data, CNNs have demonstrated their power in image restoration tasks, such as image super-resolution (SR) \cite{niu20single, yiqun21,lu22transformer}, image deblurring \cite{orest19deblurganv2, suin20spatially, cho21rethinking}, and image dehazing \cite{dong20multi, qin20ffa, wu21contrastive,li22single}. 
Several CNN-based SR techniques have been successfully utilized for medical image enhancement \cite{zhang18fast, bing19medical, mahapatra19image, lu20novel, chen21super}. 
However, the SR method alone is not sufficient for solving practical problems with highly hazy/blurry images, such as those shown in \figref{fig:fig_1}; other improvement methods must be used in conjunction.  

Other approaches for medical image dehazing/deblurring with fully supervised CNN models have been proposed \cite{liu19ganredl, armanious20medgan, shen20modeling, luthra21eformer, cheng21prior, huang22edge, sharif22deep}. 
Despite achieving attractive results, a large amount of paired LQ/HQ images for training are still indispensable and very difficult to obtain in reality. 
To solve this problem, several medical image enhancement methods based on CycleGAN \cite{zhu2017unpaired}, which do not require paired training data, have been studied. 
For retinal data, Zhao \emph{et al.} \cite{zhao19data} extended the CycleGAN and built a system for deblurring retinal images. 
Luo \emph{et al.} \cite{luo20dehaze} applied a CycleGAN variant to generate synthesized cataract-like images from clear images. 
They then trained a supervised CNN model based on the synthesized cataract-like and clear images for cataractous haze removal. 
Wan \emph{et al.} \cite{wan22rentinal} combined CycleGAN with convolutional block attention modules \cite{woo18cbam} for unsupervised retinal image enhancement. 
Chen \emph{et al.} \cite{chen22novel} used U-Net \cite{ronneberger15unet} as a backbone of CycleGAN and introduced an auxiliary loss function based on a pretrained fundus image quality classiﬁcation network to guide the system in producing high-quality images. 
Ma \emph{et al.} \cite{ma21structure} proposed a method called StillGAN with two novel constraints, namely, illumination regularization and structure loss, for general medical image enhancement and supporting clinical interpretation. 

\subsection{Medical image quality assessment}
Image quality assessment (IQA) methods play an important role in assessing the quality of medical images. 
IQA for medical images can be divided into two main categories, namely, subjective quality assessment, which is evaluated by humans, and objective quality assessment, which is evaluated by computer algorithms. 
Objective quality assessment can be further categorized into full-reference IQA (FR-IQA), where HQ reference images are available for testing, and NR-IQA, where reference images are unavailable \cite{chow16review}. 
As mentioned above, the originality aspect, particularly the structure preservation, should also be considered along with the quality aspect when evaluating medical image enhancement methods. 
Although subjective-quality assessment methods are important measurements for medical evaluations, they are very time consuming and expensive since obtaining results requires knowledge and effort from doctors. 
In addition, this evaluation method is sometimes unstable since it heavily depends on the physical conditions and emotional state of the observers.  

The FR-IQA methods can be used to evaluate not only the quality aspect but also the originality aspect. 
To measure the structure preservation of the outputs, the SSIM is commonly used among the FR-IQA methods \cite{subramani19fuzzy, liu19ganredl, luo20dehaze, armanious20medgan, shen20modeling, luthra21eformer, cheng21prior, huang22edge, sharif22deep}. 
However, as mentioned earlier, it is very difficult to obtain HQ medical reference images in practice. 
It should be noted that in cases where no reference images are available, the SSIM can still be used to evaluate the structure preservation to some degree \cite{rundo19medga, kandhway20novel, acharya21, zhang22medical}, but would produce inaccurate results if the input image is too noisy.  

The NR-IQA methods, on the other hand, do not require reference images and have been applied in several medical image enhancement studies due to their practicality \cite{zhao19data, wan22rentinal, ma21structure}. 
However, to the best of our knowledge, a remaining limitation of these NR-IQA methods is that they can only be used to measure visual quality but not structure preservation. 

\section{Proposed Methods}
    \subsection{The Laplacian structural similarity index measurement – LaSSIM}
The LaSSIM is an objective evaluation of structural preservation for medical image enhancement tasks that do not require reference images. 
LaSSIM calculates the SSIM after applying the LP process on both input and output images. 
The key idea behind LaSSIM is the ability of the LP process to effectively express the global structure of the images under conditions of high image degradation. 
\figref{fig:fig_2} shows an example of extracting different LP levels from a clean throat image and its blurred version. 
In the pixel space (level 0), the SSIM is low due to its vulnerability to blurriness. 
However, at a certain level of LP space (i.e., level 3), the SSIM between the two images are almost identical.  

Given an image $I_l$, at the first level of LP, it produces a down-sampled image $I_{(l+1)}=\mathrm{down}(I_l)$, where $\mathrm{down}(\cdot)$ is a $2\times$ down-sampling with a low-pass Gaussian filter. 
To precisely reconstruct the $I_l$ image from $I_{(l+1)}$, LP records a high-frequency residual $h_l=I_l-up(I_{(l+1)})$, where $\mathrm{up}(\cdot)$ denotes the $2\times$ up-sampling function. 
The same iterative process can be applied from the image $I_{(l+1)}$ to further reduce the image resolution. 
An $L$-level LP representation includes the original image $I_0$ at the first level, followed by the residual signals $\{h_0, h_1,...,h_{L-1}\}$. 
In this work, the LaSSIM calculates the SSIM metric on the same level residual obtained from both images $I$ and $\widehat I$. 
Concretely, LaSSIM is defined as follows: 
%% Eq. 1
\begin{equation}
    \mathrm{LaSSIM}_l(I,\widehat I)=\mathrm{SSIM}(\mathrm{LP}_l(I), \mathrm{LP}_l(\widehat I)), 
\label{eq:eq_1}
\end{equation}
where $\mathrm{LP}_l(\cdot)$ denotes the obtained residual signal $h$ of LP at level $l$.  

The LaSSIM, like SSIM, takes values from 0 to 1, with the best value being 1 when they are exactly the same in the LP space. 
In general, we suggest that the user choose a level $l$ (or a range of $l$) according to the target problem. 
Although LaSSIM is a relative evaluation, as will be demonstrated in the results section, it is a practical and robust tool for measuring structural changes for real-world non-reference medical image enhancement tasks. 
% FIGURE 3
\begin{figure}[!t]
\centering
\includegraphics[width=0.99\linewidth]{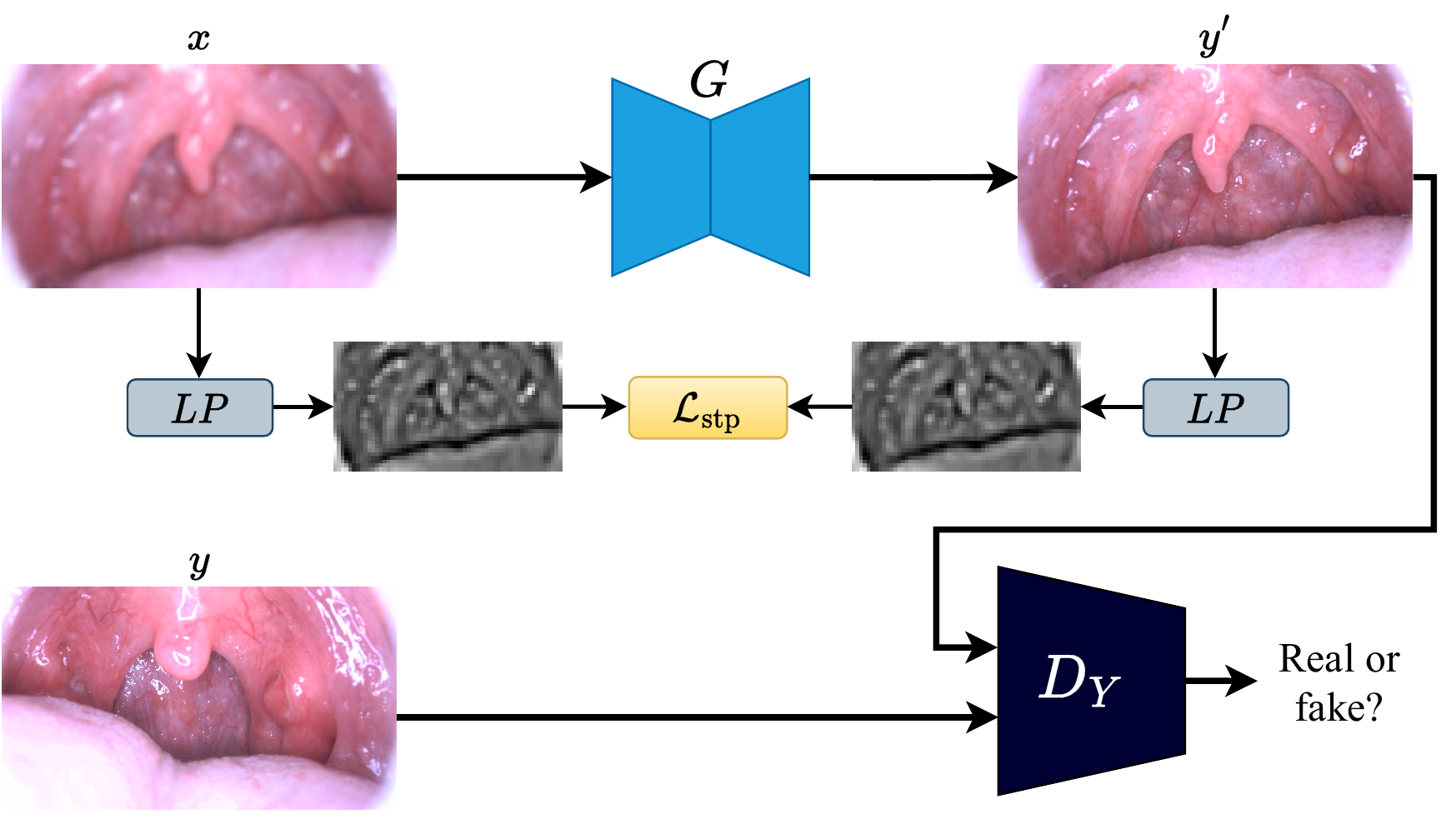}
\caption{
    Data flow of the translation from the LQ to HQ domain.
}
\label{fig:fig_3}
\end{figure}

\subsection{The unsupervised medical image enhancement method – LaMEGAN}
The LaMEGAN is an unsupervised medical image enhancement method specially designed to generate HQ-enhanced images while preserving the structure information of the inputs, supporting clinicians in medical detection and diagnosis. 
Our LaMEGAN is based on the CycleGAN, which consists of a mapping function $G:X{\rightarrow}Y$ that transforms the image from the LQ domain $X$ to the HQ domain Y and an invert mapping $F:Y{\rightarrow}X$. 
Two adversarial discriminators, $D_X$ and $D_Y$, are required for training, where $D_X$ discriminates the real LQ image $x \in X$ from the generated LQ image $x'=F(y)$ with $y \in Y$, and $D_Y$ distinguishes the real HQ image $y$ from the generated HQ image $y'=G(x)$. 
The translation $X{\rightarrow}Y$ and $Y{\rightarrow}X$ is symmetrical. 
\figref{fig:fig_3} shows the data flow of the translation from the LQ to the HQ domain.  

\subsubsection{Loss function overview}
The overall objective function $\mathcal{L}$ of LaMEGAN is also designed based on the CycleGAN with 
%% Eq. 2
\begin{equation}
    \mathcal{L}=\mathcal{L}_\mathrm{CycleGAN} + \lambda \mathcal{L}_\mathrm{stp}(G,F),
\label{eq:eq_2}
\end{equation}
where $\lambda$ is the loss coefficient and $\mathcal{L}_\mathrm{CycleGAN}$ is the original objective function from CycleGAN. 
$\mathcal{L}_\mathrm{stp}$ is the proposed structure-preserving loss term to preserve the structure information. 
Extended from our previous work MIINet \cite{cap21miinet}, we propose $\mathcal{L}_\mathrm{stp}$ to be the perceptual loss \cite{johnson16perceptual} in the LP space. 
This constraint aims to encourage natural and perceptually pleasing results while preserving the structure information from the inputs. 
The details of this loss term will be described in depth in the next section.  

CycleGAN has the following objective:
% % Eq. 3
\begin{multline}
    \mathcal{L}_\mathrm{CycleGAN} = \mathcal{L}_\mathrm{adv}(G, D_Y) + \mathcal{L}_\mathrm{adv}(F, D_X)\\
    + \beta \mathcal{L}_\mathrm{cyc}(G, F), 
\label{eq:eq_3}
\end{multline}

where $\beta$ is the cycle-consistency loss coefficient and $\mathcal{L}_\mathrm{adv}(G, D_Y)$ and $\mathcal{L}_\mathrm{adv}(F, D_X)$ are the adversarial losses for both mappings $G:X{\rightarrow}Y$ and $F:Y{\rightarrow}X$, respectively, with
% % Eq. 4
\begin{multline}
    \mathcal{L}_\mathrm{adv}(G,D_Y)=\mathbb{E}_{y{\sim}p_\mathrm{data}(y)}[(D_Y(y)-1)^2]\\
    + \mathbb{E}_{x{\sim}p_\mathrm{data}(x)}[(D_Y(y'))^2]
\label{eq:eq_4}
\end{multline}
and
% % Eq. 5
\begin{multline}
    \mathcal{L}_\mathrm{adv}(F,D_X)=\mathbb{E}_{x{\sim}p_\mathrm{data}(x)}[(D_X(x)-1)^2]\\
    + \mathbb{E}_{y{\sim}p_\mathrm{data}(y)}[(D_X(x'))^2].
\label{eq:eq_5}
\end{multline}
The cycle consistency loss $\mathcal{L}_\mathrm{cyc}(G, F)$ is defined as 
% % Eq. 6
\begin{multline}
    \mathcal{L}_\mathrm{cyc}(G,F)=\mathbb{E}_{x{\sim}p_\mathrm{data}(x)}[|F(G(x))-x|_1]\\
    +\mathbb{E}_{y{\sim}p_\mathrm{data}(y)}[|G(F(y))-y|_1].
\label{eq:eq_6}
\end{multline}

\subsubsection{The structure-preserving loss \texorpdfstring{$\mathcal{L}_\mathrm{stp}$}{}}
As mentioned above, structure preservation is important for medical image evaluation. 
In our previous study \cite{cap21miinet}, we used perceptual loss \cite{johnson16perceptual} in the pixel space to preserve the structural and color information from the inputs. 
However, some studies have shown that applying the perceptual loss directly in the pixel space leads to random unpleasant artifacts \cite{wang18esrgan, lucas19generative, tej20enhancing,amir21understanding} (which will also be demonstrated in the Results section).  

One could argue the use of SSIM or LaSSIM as an objective function to improve the structure-preserving performance of the enhancement model. 
However, using SSIM-based loss as a loss function for deep learning can lead the neural network training in the wrong direction \cite{nilsson20understandssim}. 
Our preliminary experiments also showed that simply optimizing the LaSSIM loss results in lower-quality outputs, with the color being significantly different from the inputs.  

To leverage the model to generate more natural and pleasing results while preserving the structural information from the inputs, we proposed to apply the perceptual loss to the obtained residual $h_l$ of LP on both input and output images. 
Concretely, the VGG16 model \cite{simonyan15}, which was pre-trained on the ImageNet dataset \cite{deng09imagenet}, was used for feature extraction on the obtained LP images. 
In this process, we decided to use the features obtained after the $2^{nd}$ pooling layer from the input because it preserves a better structure of the image compared to the subsequent layers used in semantic feature extraction in machine learning \cite{johnson16perceptual}. 
$\mathcal{L}_\mathrm{stp}$ is defined as
% % Eq. 7
\begin{multline}
    \mathcal{L}_\mathrm{stp}(G,F)=\mathbb{E}_{x{\sim}p_\mathrm{data}(x)}\left[\left|\phi(\mathrm{LP}_{l}(G(x))) - \phi(\mathrm{LP}_{l}(x))\right|_1\right]\\
    +\mathbb{E}_{y{\sim}p_\mathrm{data}(y)}\left[\left|\phi(\mathrm{LP}_{l}(F(y))) - \phi(\mathrm{LP}_{l}(y))\right|_1\right],
\label{eq:eq_7}
\end{multline}
where $\phi(\cdot)$ represents the features extracted from the VGG16 model. In our experiments, we chose the value $l=3$. 

\section{Experimental Setup}
    \subsection{Data collection}
In this work, we collected a large number of throat images from over 2,000 patients, including both clean and ill-conditioned images. 
These images were acquired using a camera specifically designed for taking pharyngeal images. 
Among these images, we manually inspected and selected 12,000 images with severe haze, that are out of focus, and are overexposed as LQ images and referred to them as the \say{LQ Throat} dataset. 
Another 12,000 clean and HQ images were also collected and referred to as the \say{HQ Throat} dataset (see \figref{fig:fig_1}, right). 
To evaluate the performance of the medical image enhancement models, we additionally collected 900 LQ images from 70 patients and referred to them as the test dataset. 
Note that the 70 patients were different from those used for the above training. 

\subsection{Validity assessment of the LaSSIM metric}
As mentioned above, unsupervised GAN-based methods often alter the structure of the image, which is a serious problem in medical applications. 
In this section, we investigated the validity of an image evaluation metric that can adequately capture these changes. 
Let $f(I,I')$ be an arbitrary measure that evaluates the similarity between images $I$ and $I'$. 
In addition, let $\mathscr{S}_{\mathrm{GT}}$ be a set of clean ground-truth images and its two modified versions: one with highly blurred images $\mathscr{S}_\mathrm{blur}$, and one with deformed structure and highly blurred images $\mathscr{S}_\mathrm{deform+blur}$. 
Finally, $\mathbb{F}(\mathscr{S},\mathscr{S}')$ is defined as the normalized probability distribution of the $f$ scores obtained from each image in the dataset $\mathscr{S}$ and $\mathscr{S}'$.  

If $f$ is the desired metric that can capture structural changes, then for each image $s_\mathrm{GT}$, $s_\mathrm{blur}$, $s_\mathrm{deform+blur}$ taken from each of the above datasets, $f(s_\mathrm{GT},s_\mathrm{blur})>f(s_\mathrm{GT},s_\mathrm{deform+blur})$ should hold. 
That is, its distribution $\mathbb{F}(\mathscr{S}_\mathrm{GT},\mathscr{S}_\mathrm{blur})>\mathbb{F}(\mathscr{S}_\mathrm{GT},\mathscr{S}_\mathrm{deform+blur})$ is also valid. 
In this experiment, both the SSIM and LaSSIM (with the extracted residual level set to $l=3$) metrics were validated as $f$ (as well as $\mathbb{F}$). 
In addition, the distribution difference of $\mathbb{F}(\mathscr{S}_\mathrm{GT}, \mathscr{S}_\mathrm{blur})$ and $\mathbb{F}(\mathscr{S}_\mathrm{GT}, \mathscr{S}_\mathrm{deform+blur})$ for both SSIM and LaSSIM was evaluated using the Jensen-Shannon divergence.  

\begin{figure}[!t]
\centering
\includegraphics[width=0.99\linewidth]{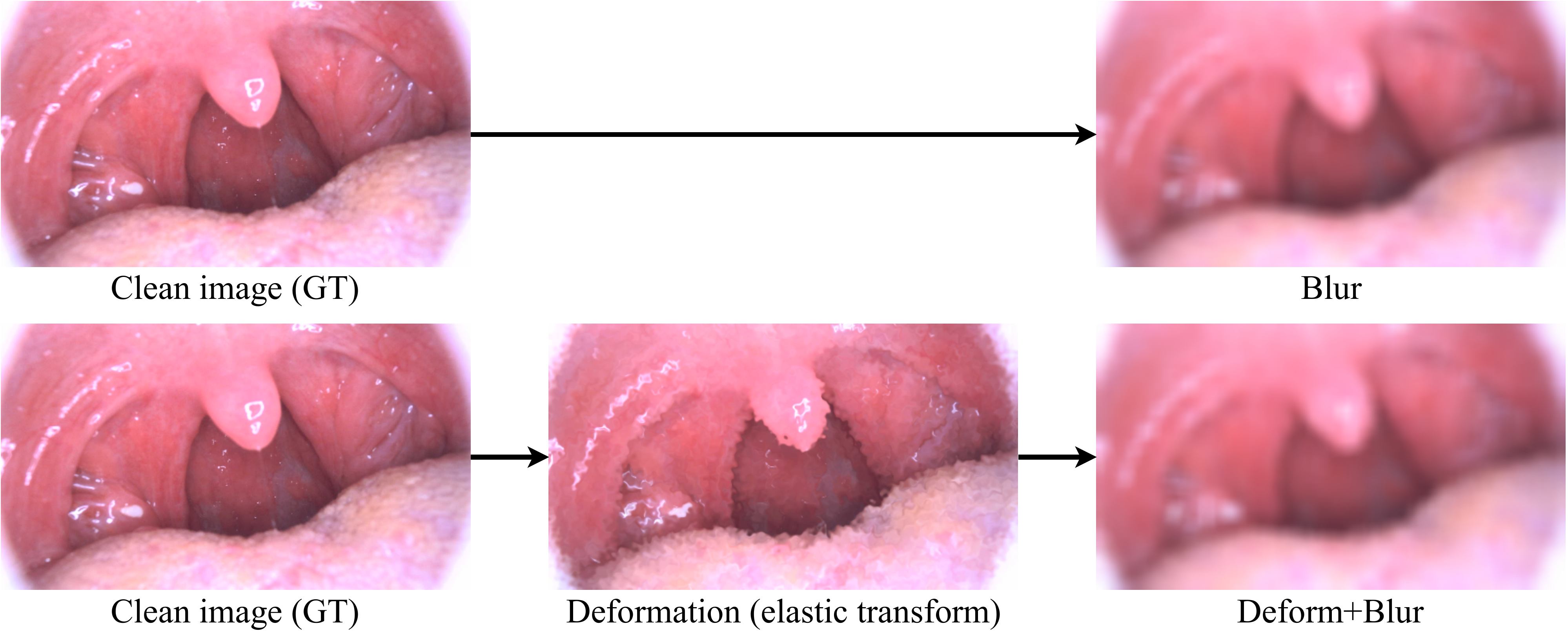}
\caption{
    Step-by-step process for obtaining the $\mathscr{S}_\mathrm{GT}$, $\mathscr{S}_\mathrm{blur}$, and $\mathscr{S}_\mathrm{deform+blur}$ image datasets for our experiments.
}
\label{fig:fig_4}
\end{figure}

For the experimental setup, the $\mathscr{S}_\mathrm{GT}$ set was first obtained by randomly selecting 1,000 clear images from the HQ Throat dataset. 
The $\mathscr{S}_\mathrm{blur}$ images were created by blurring the images in the $\mathscr{S}_\mathrm{GT}$ set. 
The $\mathscr{S}_\mathrm{deform+blur}$ images were generated with elastic transform \cite{simard03best} on $\mathscr{S}_\mathrm{GT}$, and then blur was added on top. 
Specifically for the elastic transformation, random displacement fields (i.e., random offset values $\Delta x(x,y)$, $\Delta y(x,y)$ at all pixel coordinates) were first generated. 
The fields $\Delta x$ and $\Delta y$ were then convolved with a Gaussian function with standard deviation $\sigma$ (in pixels). 
The results were multiplied by a scaling factor $\alpha$ that controls the intensity of the deformation. 
\figref{fig:fig_4} illustrates the step-by-step process of obtaining these three sets.  

With respect to the intensity of image degradation, three levels of intensity were used for each deformation (def) and blur: low (L), medium (M), and high (H). As a result, three different combinations were generated for $\mathscr{S}_\mathrm{blur}$ images and nine combinations for $\mathscr{S}_\mathrm{deform+blur}$ images, such as [$\mathrm{L_{def}}$, $\mathrm{H_{blur}}$], [$\mathrm{M_{def}}$, $\mathrm{H_{blur}}$], and so on. 
Please refer to our GitHub repository for more experimental details. 

\subsection{Evaluation methods for medical image enhancement}
We considered two aspects in evaluating medical image enhancement: quality (e.g., clearness, naturalness of the outputs) and originality (e.g., change in structure, texture, color from the inputs). 
In our previous work, we introduced the mean doctor opinion score (MDOS) \cite{cap21miinet}. 
This time, we extended it as MDOS-Q for the quality aspect and MDOS-O for the originality aspect. 
For the MDOS-Q, physicians score the images based on the question, \say{\textbf{\emph{How good is this image in terms of quality to support medical diagnosis (e.g., clearness, naturalness)?}}}. 
For the MDOS-O, the evaluation is based on the question \say{\textbf{\emph{How much originality does the output retain from the input (e.g., structure preservation, texture, color)?}}}. 
For both the MDOS-Q and MDOS-O, image scores are scaled from 1.0 (poor) to 5.0 (excellent), with higher scores being better. 
In our evaluation, we collected both scores from three clinicians. 
The details are described later.  

As objective evaluation measures, we used the proposed LaSSIM to evaluate the structure preservation and a total of seven NR-IQA criteria, including NIQE \cite{mittal12niqe}, BRISQUE \cite{mittal12brisque}, NIMA \cite{talebi18nima}, PaQ-2-PiQ \cite{ying20patches}, DBCNN \cite{zhang18dbcnn}, HyperIQA \cite{su20hyperiqa}, and MUSIQ \cite{ke21musiq}, to measure the image quality.  

In summary, from the originality point of view, our LaSSIM and MDOS-O were applied to evaluate the structure preservation and the texture and color constancy, respectively. From the quality point of view, the above seven quantitative NR-IQA criteria and the MDOS-Q subjective score were used. 
% FIGURE 4
% FIGURE 5
\begin{figure}[!t]
\centering
\includegraphics[width=0.99\linewidth]{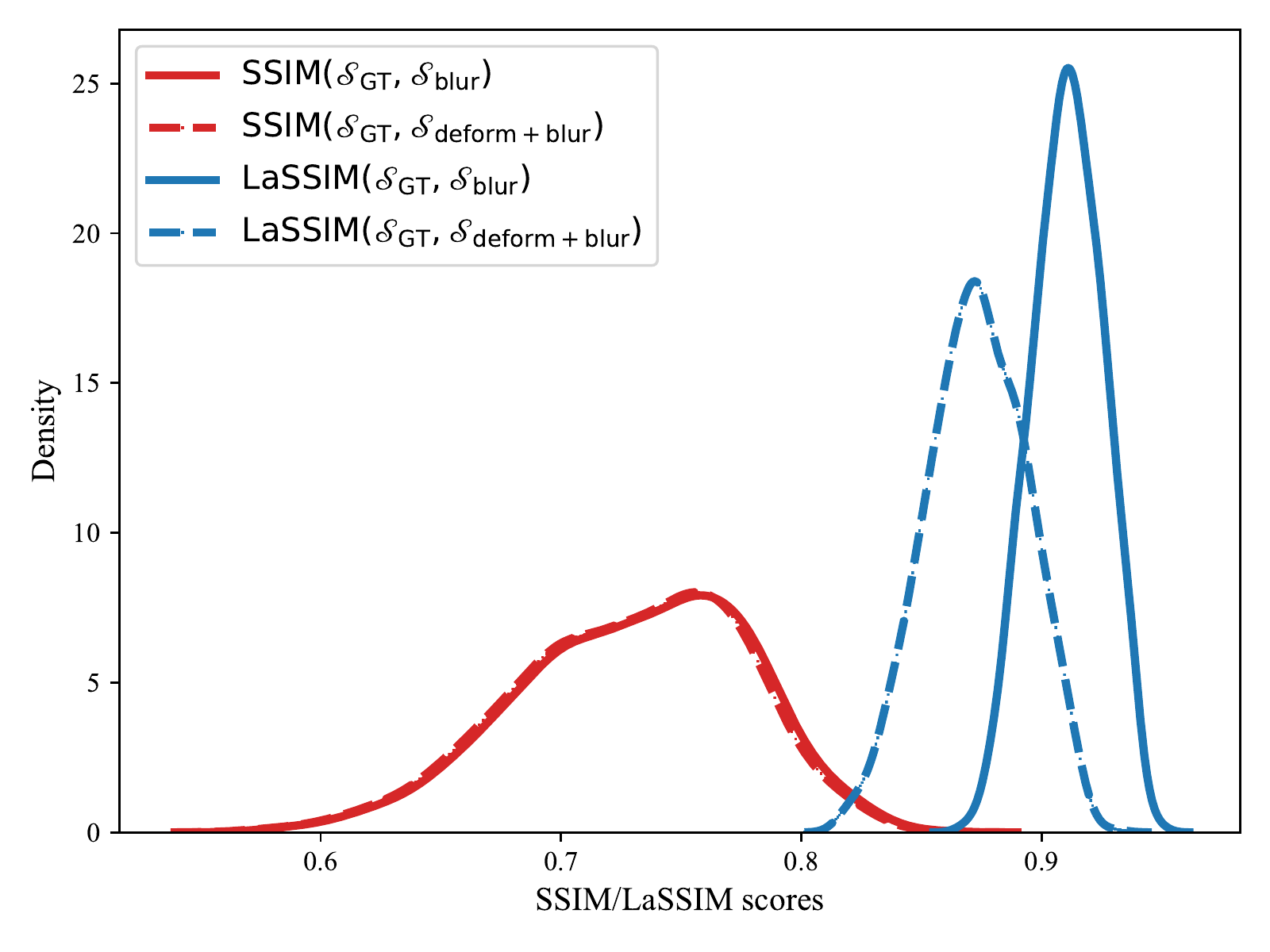}
\caption{
    Comparison of $\mathbb{F}(\mathscr{S}_\mathrm{GT},\mathscr{S}_\mathrm{blur})$ and $\mathbb{F}(\mathscr{S}_\mathrm{GT},\mathscr{S}_\mathrm{deform+blur})$ at the distortion level [$\mathrm{M_{def}}$, $\mathrm{H_{blur}}$]. Under strong blur conditions, SSIM cannot detect distortions.
}
\label{fig:fig_5}
\end{figure}

\subsection{Training the LaMEGAN}
To train our LaMEGAN, we used the same training procedures described in CycleGAN \cite{zhu2017unpaired}. 
The $\lambda$ and $\beta$ in \equref{eq:eq_2} and (\ref{eq:eq_3}) were set to 0.25 and 10.0, respectively, based on our preliminary experiments. 
Due to GPU memory limitations, the resolution of the input and output in this work was set to $480 \times 272$ pixels. 
For comparison purposes, we also trained three other medical image enhancement models using CycleGAN, MIINet \cite{cap21miinet}, and StillGAN \cite{ma21structure} with their default hyperparameters. 
The training process was completed after 200 epochs for all four models. 
For further training details, please refer to the CycleGAN literature \cite{zhu2017unpaired}. 
Note that we only used the one-way translation of $\mathrm{LQ} \rightarrow \mathrm{HQ}$ for the evaluation. 

\section{Results}
    % TABLE I
% Please add the following required packages to your document preamble:
% \usepackage{booktabs}
% \usepackage{multirow}
\begin{table*}[t!]
\centering
\caption{Structure preservation evaluation comparisons of SSIM and LaSSIM under different intensities}
\label{tab:table_1}
\resizebox{0.8\linewidth}{!}{%
\begin{tabular}{@{}ccccccc@{}}
\toprule
\multirow{2}{*}{\begin{tabular}[c]{@{}c@{}}Intensity of \\ image degradation\end{tabular}} &
  \begin{tabular}[c]{@{}c@{}}$\mathrm{SSIM}(\mathscr{S}_\mathrm{GT},$ \\ $\mathscr{S}_\mathrm{blur})$\end{tabular} &
  \begin{tabular}[c]{@{}c@{}}$\mathrm{SSIM}(\mathscr{S}_\mathrm{GT},$ \\ $\mathscr{S}_\mathrm{deform+blur})$\end{tabular} &
  \multirow{2}{*}{$JS_{\mathrm{SSIM}}$} &
  \begin{tabular}[c]{@{}c@{}}$\mathrm{LaSSIM}(\mathscr{S}_\mathrm{GT},$ \\ $\mathscr{S}_\mathrm{blur})$\end{tabular} &
  \begin{tabular}[c]{@{}c@{}}$\mathrm{LaSSIM}(\mathscr{S}_\mathrm{GT},$ \\ $\mathscr{S}_\mathrm{deform+blur})$\end{tabular} &
  \multirow{2}{*}{$JS_{\mathrm{LaSSIM}}$} \\ \cmidrule(lr){2-3} \cmidrule(lr){5-6}
                      & mean$\pm$std    & mean$\pm$std    &       & mean$\pm$std    & mean$\pm$std    &                \\ \midrule
{[$\mathrm{L_{def}}$, $\mathrm{L_{blur}}$]} & 0.816$\pm$0.036 & 0.805$\pm$0.038 & 0.105 & 0.991$\pm$0.002 & 0.970$\pm$0.006 & \textbf{0.832} \\
{[$\mathrm{M_{def}}$, $\mathrm{L_{blur}}$]} & 0.816$\pm$0.036 & 0.788$\pm$0.038 & 0.252 & 0.991$\pm$0.002 & 0.941$\pm$0.012 & \textbf{0.835} \\
{[$\mathrm{H_{def}}$, $\mathrm{L_{blur}}$]} & 0.816$\pm$0.036 & 0.766$\pm$0.040 & 0.413 & 0.991$\pm$0.002 & 0.902$\pm$0.019 & \textbf{0.839} \\
{[$\mathrm{L_{def}}$, $\mathrm{M_{blur}}$]} & 0.762$\pm$0.044 & 0.759$\pm$0.044 & 0.022 & 0.964$\pm$0.006 & 0.946$\pm$0.010 & \textbf{0.633} \\
{[$\mathrm{M_{def}}$, $\mathrm{M_{blur}}$]} & 0.762$\pm$0.044 & 0.754$\pm$0.044 & 0.065 & 0.964$\pm$0.006 & 0.918$\pm$0.015 & \textbf{0.817} \\
{[$\mathrm{H_{def}}$, $\mathrm{M_{blur}}$]} & 0.762$\pm$0.044 & 0.746$\pm$0.045 & 0.130 & 0.964$\pm$0.006 & 0.884$\pm$0.020 & \textbf{0.833} \\
{[$\mathrm{L_{def}}$, $\mathrm{H_{blur}}$]} & 0.731$\pm$0.048 & 0.730$\pm$0.047 & 0.007 & 0.911$\pm$0.015 & 0.895$\pm$0.017 & \textbf{0.320} \\
{[$\mathrm{M_{def}}$, $\mathrm{H_{blur}}$]} & 0.731$\pm$0.048 & 0.728$\pm$0.047 & 0.023 & 0.911$\pm$0.015 & 0.873$\pm$0.021 & \textbf{0.598} \\
{[$\mathrm{H_{def}}$, $\mathrm{H_{blur}}$]} & 0.731$\pm$0.048 & 0.725$\pm$0.047 & 0.047 & 0.911$\pm$0.015 & 0.843$\pm$0.026 & \textbf{0.757} \\ \bottomrule
\end{tabular}
}
\end{table*}
\subsection{Results on validity assessment of the LaSSIM}
\figref{fig:fig_5} visualizes $\mathbb{F}(\mathscr{S}_\mathrm{GT},\mathscr{S}_\mathrm{blur})$ and $\mathbb{F}(\mathscr{S}_\mathrm{GT},\mathscr{S}_\mathrm{deform+blur})$ at the distortion level [$\mathrm{M_{def}}$, $\mathrm{H_{blur}}$], where $\mathbb{F}=\{\mathrm{SSIM}$ and $\mathrm{LaSSIM}\}$. 
Despite the fact that the image structure of $\mathscr{S}_\mathrm{deform+blur}$ has been distorted from $\mathscr{S}_\mathrm{GT}$, the two score distributions of $\mathrm{SSIM}(\mathscr{S}_\mathrm{GT}, \mathscr{S}_\mathrm{blur})$ and $\mathrm{SSIM}(\mathscr{S}_\mathrm{GT}, \mathscr{S}_\mathrm{deform+blur})$ are nearly identical. 
By contrast, the score distributions of $\mathrm{LaSSIM}(\mathscr{S}_\mathrm{GT}, \mathscr{S}_\mathrm{blur})$ and $\mathrm{LaSSIM}(\mathscr{S}_\mathrm{GT}, \mathscr{S}_\mathrm{deform+blur})$ are distinctly separated, indicating that LaSSIM has a much better capability than SSIM in evaluating the underlying structural changes under strong blur conditions. 
\tabref{tab:table_1} shows the comparison of the structure preservation evaluation between SSIM and LaSSIM under nine distortion level combinations, wherein $JS_{\mathrm{SSIM}}$ indicates the Jensen-Shannon divergence between $\mathrm{SSIM}(\mathscr{S}_\mathrm{GT},\mathscr{S}_\mathrm{blur})$ and $\mathrm{SSIM}(\mathscr{S}_\mathrm{GT},\mathscr{S}_\mathrm{deform+blur})$ at each image degradation combination. 
Similarly, $JS_{\mathrm{LaSSIM}}$ indicates the LaSSIM distributions. 
In all nine distortion combinations, the $JS_{\mathrm{LaSSIM}}$ is always much larger than the $JS_{\mathrm{SSIM}}$. 
This result indicates that SSIM does not correctly detect structural degradation when the image is blurred. 
On the other hand, LaSSIM is a reliable and robust metric for evaluating structure preservation. 
We also performed similar experiments on different datasets, such as ImageNet \cite{deng09imagenet}, ChestX-ray14 \cite{wang17chestx}, CAMELYON16 \cite{CAMELYON16}, and ISIC2018 \cite{ISIC2018}, and the results showed that the LaSSIM is robust for evaluating structure preservation on these datasets. More results of the LaSSIM can be found on our GitHub page. 

\subsection{Results on throat image enhancement}
Three physicians were enrolled to score the MDOS-Q and MDOS-O in this work. 
Due to the large number of test images, 160 images were selected for each image. 
The physicians evaluated the original LQ throat image and the quality-enhanced images using the four methods, including the proposed method (i.e., CycleGAN, StillGAN, MIINet, and LaMEGAN). 
In the end, a total of 800 images were evaluated by each physician. 
The final MDOS-Q and MDOS-O for each image were averaged across all physicians.  

\begin{figure*}[!t]
\centering
\includegraphics[width=0.99\linewidth]{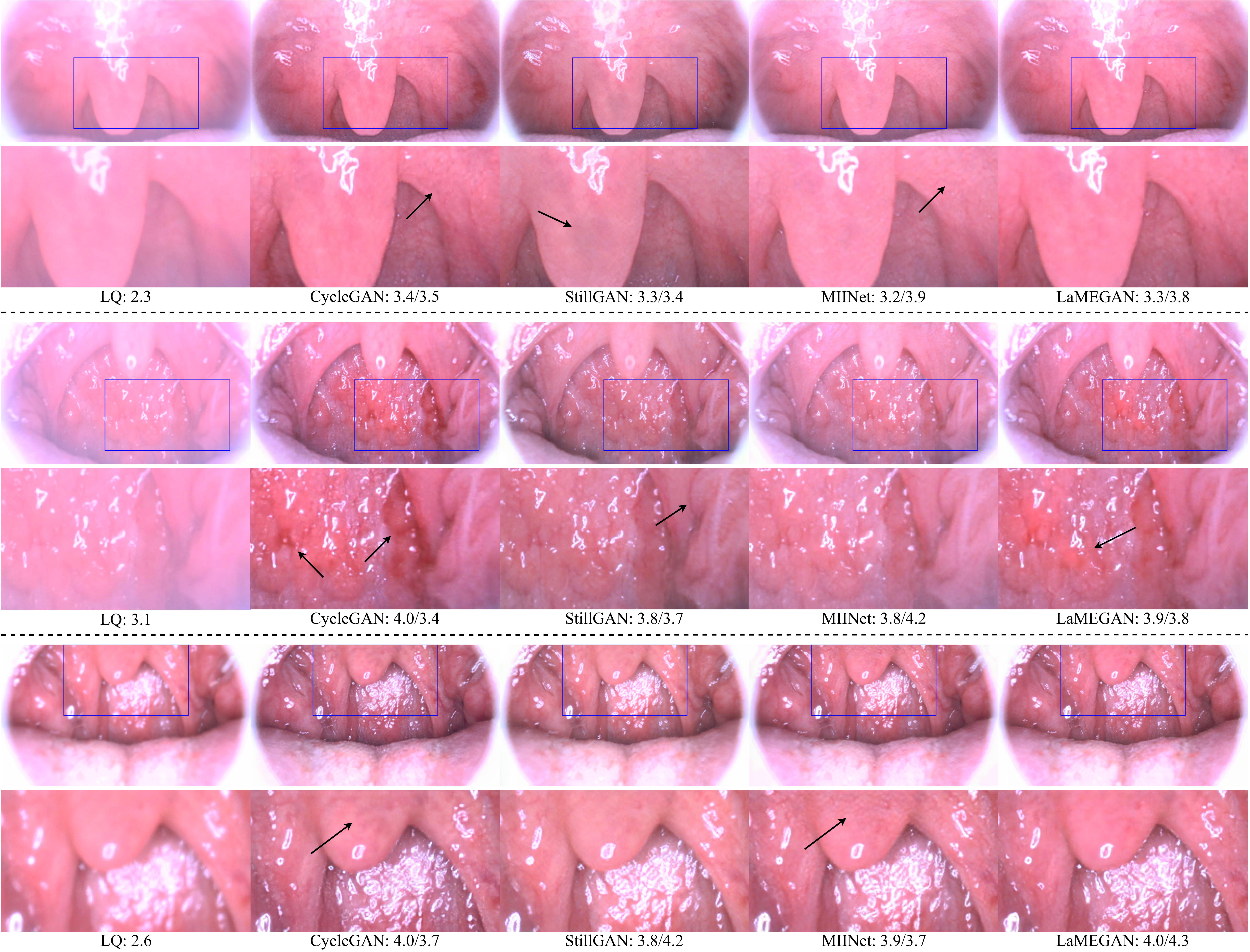}
\caption{
    Visual comparison between LQ, CycleGAN, StillGAN, MIINet, and LaMEGAN. The scores at the LQ images are the MDOS-Q, while the rest are MDOS-Q/MDOS-O. Black arrows indicate structure/texture changes or sudden changes in color from the inputs.
}
\label{fig:fig_6}
\end{figure*}

\begin{figure*}[!t]
\centering
\begin{subfigure}[b]{0.49\textwidth}
    \centering
    \includegraphics[width=\textwidth]{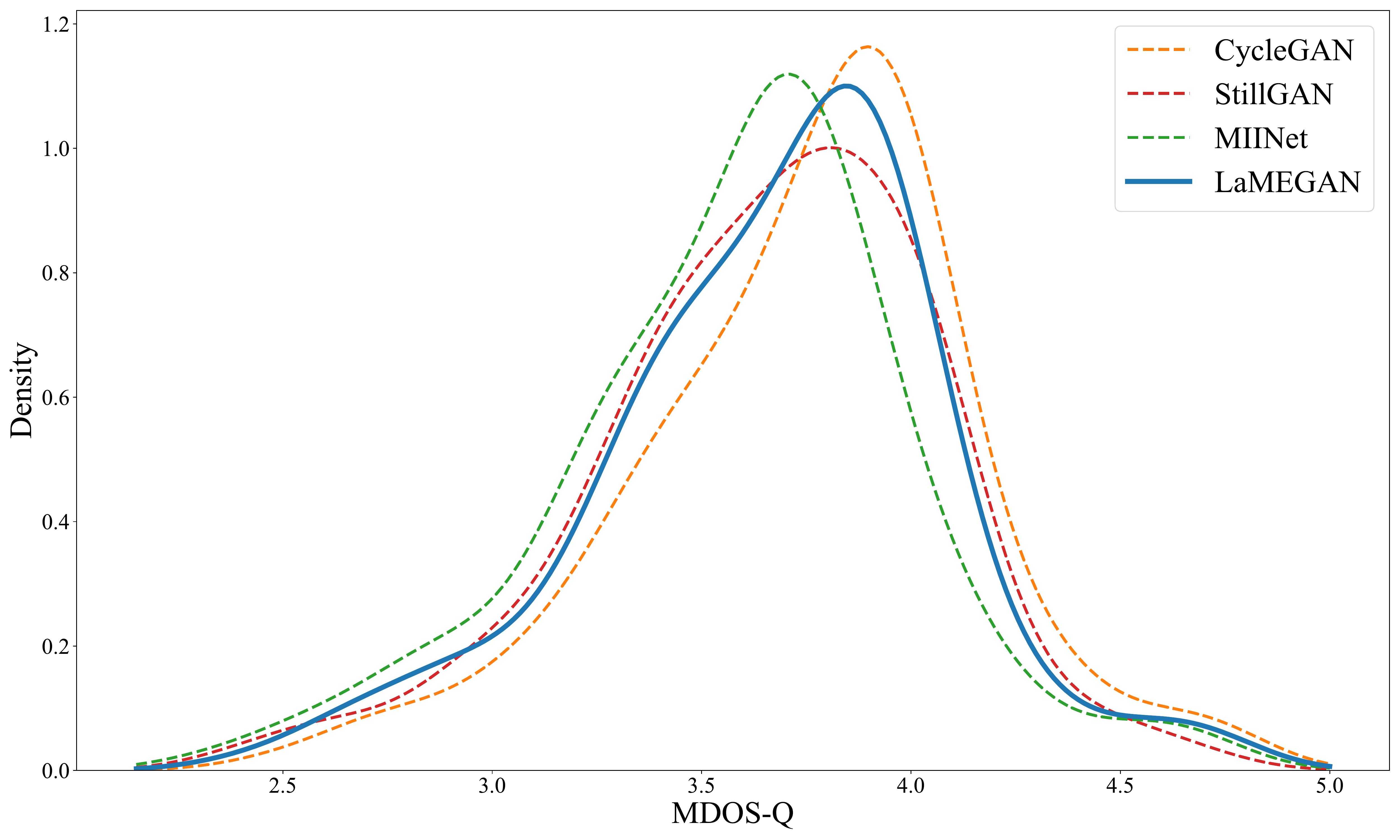}
\end{subfigure}
\hspace{0em}
\begin{subfigure}[b]{0.49\textwidth}
    \centering
    \includegraphics[width=\textwidth]{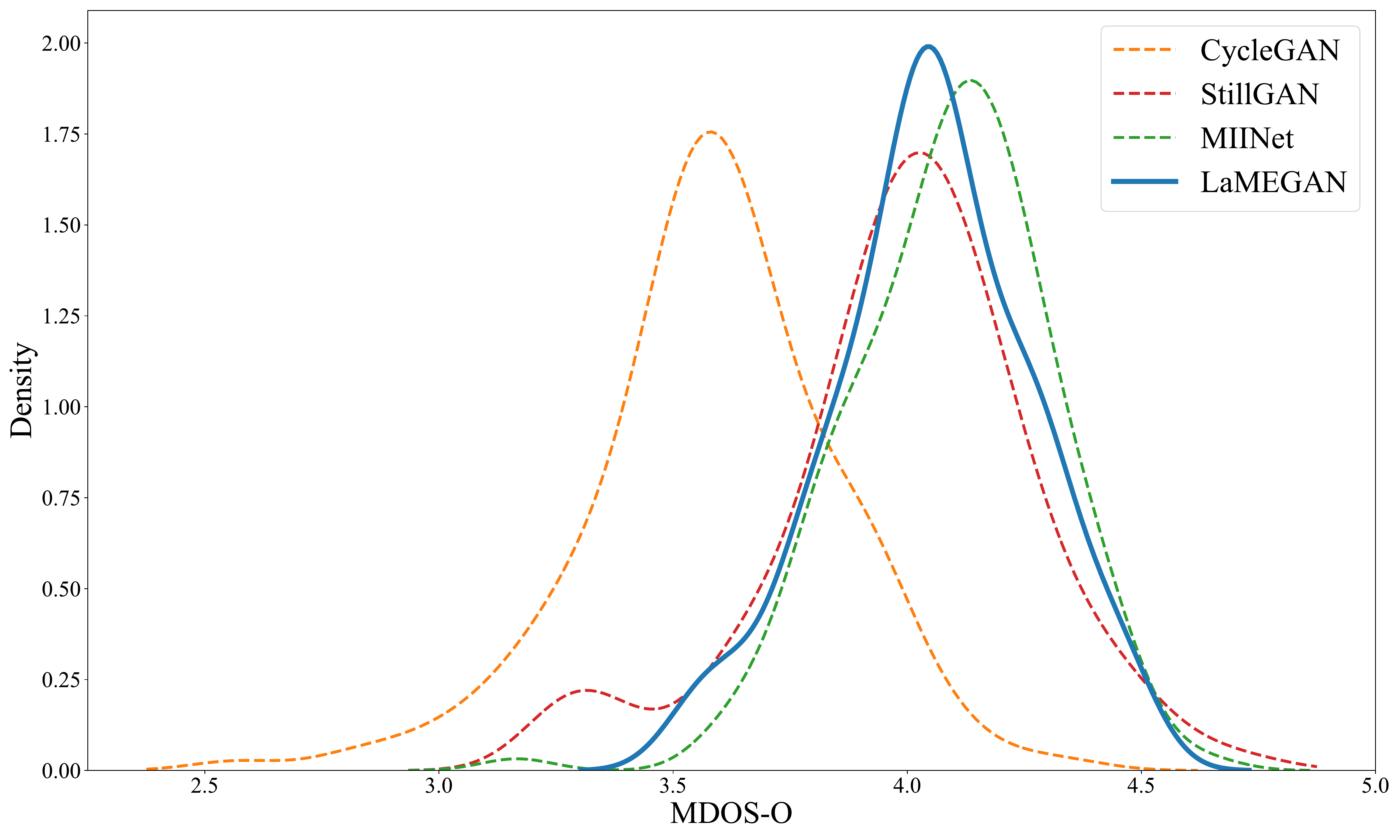}
\end{subfigure}
\caption{
    Distributions of the MDOS-Q and MDOS-O evaluation based on 160 images. While other methods are affected by the quality-originality trade-off, the proposed LaMEGAN achieves a satisfactory balance between MDOS-Q and MDOS-O.
}
\label{fig:fig_7}
\end{figure*}

Visual comparisons of the original LQ images generated from CycleGAN, StillGAN, MIINet, and LaMEGAN are shown in \figref{fig:fig_6}. 
While the CycleGAN, StillGAN, and MIINet methods changed the structure/texture of input images, our LaMEGAN method not only exhibited robustness in preserving the original attributes but also effectively enhanced the quality from the LQ inputs. 
Tables \ref{tab:table_2} and \ref{tab:table_3} show the numerical results of the image enhancement in the quality and originality aspects. 
The text in \textcolor{red}{\textbf{red}} and \textcolor{black}{\textbf{black}} indicate the best and the second-best results, respectively. 
\figref{fig:fig_7} shows the distributions of both MDOS-Q and MDOS-O scores. 
Although CycleGAN showed the best results in all categories in terms of image quality, it was very limited in terms of structural reproducibility. 
On the other hand, the proposed LaMEGAN achieved image quality scores close to CycleGAN, while its results in structural reproducibility were on par with the best model MIINet, which has inferior image quality. 

\section{Discussion}
    % FIGURE 6
% Table II
\begin{table*}[t!]
\centering
\caption{Quality aspect results from seven NR-IQA methods and the subjective MDOS-Q metric}
\label{tab:table_2}
\resizebox{0.95\linewidth}{!}{%
\begin{threeparttable}
\begin{tabular}{lcccccccc}
\hline
\multicolumn{1}{c}{} & NIQE ($\downarrow$)   & BRISQUE ($\downarrow$)  & NIMA ($\uparrow$)   & PaQ-2-PiQ ($\uparrow$)        & DBCNN ($\uparrow$)   & MUSIQ ($\uparrow$)   & HyperIQA ($\uparrow$) & MDOS-Q* ($\uparrow$) \\ \hline
LQ                   & 7.19$\pm$1.28 & 38.54$\pm$10.19 & 3.67$\pm$0.35 & 69.56$\pm$4.54          & 41.26$\pm$8.83 & 42.10$\pm$8.31 & 39.84$\pm$5.72  & 2.86$\pm$0.32  \\
CycleGAN \cite{zhu2017unpaired} &
  {\color[HTML]{FF0000} \textbf{4.27$\pm$0.53}} &
  {\color[HTML]{FF0000} \textbf{21.71$\pm$6.39}} &
  {\color[HTML]{FF0000} \textbf{4.11$\pm$0.26}} &
  {\color[HTML]{FF0000} \textbf{75.30$\pm$1.51}} &
  {\color[HTML]{FF0000} \textbf{59.10$\pm$5.15}} &
  {\color[HTML]{FF0000} \textbf{57.24$\pm$5.02}} &
  {\color[HTML]{FF0000} \textbf{54.39$\pm$5.08}} &
  {\color[HTML]{FF0000} \textbf{3.75$\pm$0.16}} \\
StillGAN \cite{ma21structure}             & 4.98$\pm$0.68 & 24.09$\pm$5.61  & 4.04$\pm$0.27 & \textbf{75.01$\pm$1.86} & 57.80$\pm$5.57 & 56.23$\pm$5.81 & 52.67$\pm$5.00  & 3.66$\pm$0.19  \\
MIINet \cite{cap21miinet}               & 4.50$\pm$0.63 & 22.09$\pm$5.85  & 3.98$\pm$0.25 & 74.29$\pm$1.78          & 56.73$\pm$5.40 & 54.54$\pm$6.04 & 51.02$\pm$5.35  & 3.58$\pm$0.21  \\
\begin{tabular}[c]{@{}l@{}}LaMEGAN\\ (proposed)\end{tabular} &
  \textbf{4.45$\pm$0.53} &
  \textbf{21.96$\pm$5.54} &
  \textbf{4.05$\pm$0.25} &
  74.91$\pm$1.63 &
  \textbf{58.03$\pm$4.99} &
  \textbf{56.36$\pm$5.39} &
  \textbf{53.39$\pm$5.34} &
  \textbf{3.67$\pm$0.17} \\ \hline
\end{tabular}
\begin{tablenotes}
    \item[$\ast$] {Evaluated on 160 images}
\end{tablenotes}
\end{threeparttable}
}
\end{table*}
% FIGURE 7
% Table III
\begin{table}[t!]
\centering
\caption{Originality aspect results from LaSSIM and the subjective MDOS-O metric}
\label{tab:table_3}
\resizebox{0.69\linewidth}{!}{%
\begin{threeparttable}
\begin{tabular}{@{}lcc@{}}
\toprule
\multicolumn{1}{c}{} & LaSSIM ($\uparrow$)   & MDOS-O* ($\uparrow$)         \\ \midrule
LQ                                                           & {\color[HTML]{FF0000} \textbf{1.0}} & {\color[HTML]{FF0000} \textbf{5.0}} \\
CycleGAN \cite{zhu2017unpaired}             & 0.869$\pm$0.049 & 3.58$\pm$0.32          \\
StillGAN \cite{ma21structure}             & 0.895$\pm$0.069 & 3.99$\pm$0.29          \\
MIINet \cite{cap21miinet}               & 0.925$\pm$0.039 & \textbf{4.09$\pm$0.09} \\
\begin{tabular}[c]{@{}l@{}}LaMEGAN\\ (proposed)\end{tabular} & \textbf{0.936$\pm$0.038}                & 4.05$\pm$0.12                           \\ \bottomrule
\end{tabular}
\begin{tablenotes}
    \item[$\ast$] {Evaluated on 160 images}
\end{tablenotes}
\end{threeparttable}
}
\end{table}

\subsection{The effectiveness of the LaSSIM}
The LaSSIM metric is designed for evaluating the structure preservation of unsupervised medical image enhancement tasks without the need for reference images. 
The results from \figref{fig:fig_5} and Table \ref{tab:table_1} show the superior structure preservation evaluation performance of LaSSIM compared with SSIM under different blurry conditions. 
The SSIM is a pixel-based metric and therefore, strong blur reduces its ability to evaluate structural changes. 
Owing to its ability to express the underlying structure of the LPs, our LaSSIM has shown its practicality in capturing structure changes not only on our endoscopy throat data but also on a wide range of image datasets.  

Since LaSSIM is an unsupervised evaluation metric, the scores obtained are environment-dependent. 
In other words, it should be used for relative comparisons between methods in a given environment and not the absolute score obtained in an evaluation based on supervised methods.

\subsection{The quality-originality trade-off of medical image enhancement models}
We have studied the performance of several unsupervised medical image enhancement methods on various evaluation metrics. 
The quality-originality trade-off can be clearly observed in the case of the CycleGAN model. 
Although it achieved the highest quality scores in all seven NR-IQA and the MDOS-Q metrics (Table \ref{tab:table_2}), the CycleGAN obtained the lowest LaSSIM and MDOS-O among all methods (Table \ref{tab:table_3}). 
As visually shown in \figref{fig:fig_6}, despite generating very clear images, CycleGAN either notably changed the color/structure from the inputs or randomly generated artifacts. 
Since CycleGAN prioritizes learning the distribution of the target data over preserving the original structure, it has greater flexibility in generating images that are close to the target HQ domain but also loses the ability to maintain originality from the inputs.  

StillGAN, as one of the state-of-the-art methods for medical image enhancement, generates relatively HQ images (MDOS-Q and NF-IQA) but cannot retain the original color and often turns some regions of the input images into much darker ones. 
Thus, StillGAN achieved a comparatively low result on LaSSIM, as reported in Table \ref{tab:table_3}. 
This result is due to the fact that LaSSIM is based on SSIM, and it is known that SSIM is affected by local color/luminance changes \cite{wang04ssim, qureshi12}. 
We suspect that the high diversity observed within our throat dataset may have contributed to the inconsistent color outputs produced by StillGAN. 
This issue was also mentioned in their paper as the model has some limitations in handling complex datasets. 
Note that the LaSSIM of StillGAN can be improved by increasing the coefficient weight of their SSIM-based loss term. 
However, too large a value of this weight usually results in amplifying noise or producing artifacts \cite{ma21structure}.  

Through the utilization of perceptual loss, both LaMEGAN and MIINet achieved better color preservation than StillGAN. 
However, since MIINet optimized the perceptual loss in the pixel space, it generated more random textures and artifacts (see \figref{fig:fig_6}, first and last examples). 
Applying perceptual loss in the pixel space also limited the dehazing ability as MIINet generated notably hazier images (\figref{fig:fig_6}, second example). 
Thus, MIINet obtained very high originality scores, but the image quality results were deficient. 
Taking the quality-originality trade-off into consideration, our LaMEGAN achieved a satisfactory balance in both quality and originality aspects (\figref{fig:fig_7}). 
By optimizing the perceptual loss in the LP space, the LaMEGAN model was implicitly guided to preserve the structural information. 
As a result, unwanted artifacts and random textures were effectively reduced, producing visually appealing images (\figref{fig:fig_6}). 
Moreover, the LaMEGAN achieved close quality scores from all seven NR-IQA compared with CycleGAN and attained the highest LaSSIM score among all methods, as shown in Table \ref{tab:table_2} and \ref{tab:table_3}.  

Although it achieved promising results, the LaMEGAN occasionally produced bold red areas in the generated images, which can potentially result in misdiagnosis (\figref{fig:fig_6}, second example). 
This drawback can be attributed to the complexities present in our real-world dataset, and we believe that introducing a loss term for color preservation can mitigate this issue. 
We aim to further enhance our system by addressing these current limitations in our future works. 

\section{Conclusion}
    In this study, we introduced two proposals as a framework for practical unsupervised medical image enhancement. 
First, we proposed the LaSSIM, an objective structure preservation evaluation metric for real-world medical image enhancement tasks. 
Our LaSSIM did not require any reference images and showed robustness in capturing structural changes under different degradation levels. 
Second, we proposed the LaMEGAN as a medical image enhancement method for supporting medical decision-making. 
By applying the perceptual loss in the LP extracted space, our LaMEGAN proved to be effective in not only generating pleasing enhanced results but also having the ability to preserve the originality of the LQ medical input images. 
We strongly believe that these contributions are useful tools for physicians and can have a positive impact on the field of medical imaging. 
    
% conference papers do not normally have an appendix
% use section* for acknowledgment
\section*{Acknowledgment}
We would like to thank all researchers at Aillis Inc., especially all the AI Development team members, Dr. Memori Fukuda (MD), Dr. Kei Katsuno (MD, MPH), and Dr. Takaya Hanawa (MD) for their valuable comments and feedback.

\nocite{*}
\footnotesize{
\bibliographystyle{IEEEtran}
\bibliography{reference}
}

% that's all folks
\end{document}